\documentclass[journal=nalefd,manuscript=letter]{achemso}

\usepackage{graphicx}
\usepackage[utf8]{inputenc}
\usepackage{amsmath,amsfonts,amssymb}
\usepackage{color}

\title{Atomic Force Extrema Induced by \\ the Bending of a CO-Functionalized Probe}

\author{Nicolas Néel}
\email{nicolas.neel@tu-ilmenau.de, fax:+49-3677-693602}
\affiliation{Institut für Physik, Technische Universität Ilmenau, D-98693 Ilmenau, Germany}

\author{Jörg Kröger}
\affiliation{Institut für Physik, Technische Universität Ilmenau, D-98693 Ilmenau, Germany}

\begin{document}

\maketitle

\newpage

\begin{abstract}
The control and observation of reactants forming a chemical bond at the single-molecule level is a longstanding challenge in quantum physics and chemistry.
Using a single CO molecule adsorbed at the apex of an atomic force microscope tip together with a Cu(111) surface, the molecular bending is induced by torques due to van der Waals attraction and Pauli repulsion.
As a result, the vertical force exhibits a characteristic dip-hump evolution with the molecule--surface separation, which depends sensitively on the initial tilt angle the CO axis encloses with the surface normal.
The experimental force data are reproduced by model calculations that consider the CO deflection in a harmonic potential and the molecular orientation in the Pauli repulsion term of the Lennard-Jones potential. 
The presented findings shed new light on vertical-force extrema that can occur in scanning probe experiments with functionalized tips.

\begin{tocentry}
\begin{center}
\includegraphics[width=\linewidth]{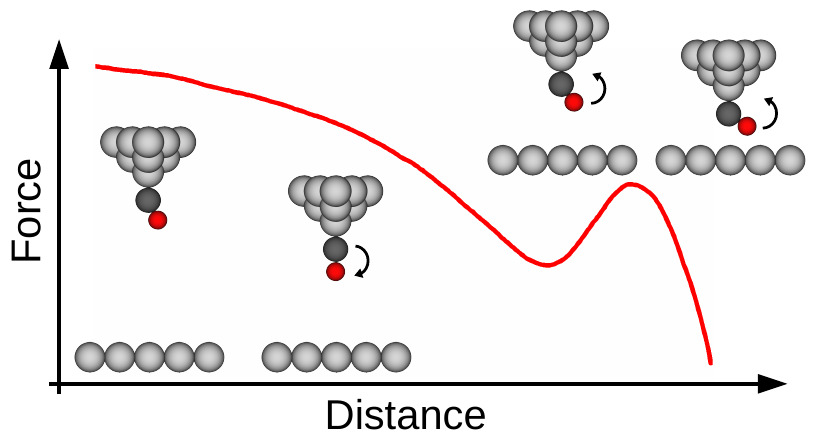}
\label{For TOC only}
\end{center}
\end{tocentry}

\end{abstract}


The carbon monoxide (CO) molecule has been serving as an archetypical molecule in surface chemistry to study adsorption processes \cite{jpc_68_2772,prl_85_3309,prb_76_195440,pccp_20_25892}.
It has attained tremendously renewed interest when astounding submolecular resolution was achieved in atomic force microscope (AFM) experiments using a tip terminated by a single CO molecule \cite{science_325_1110}.
Several seminal works using CO-functionalized probes followed and reported, e.\,g., bond order discrimination \cite{science_337_1326} and the chemical reaction of molecules on surfaces \cite{nl_14_2251,jacs_137_7424}. 
In order to explain the contrast mechanism in general \cite{prb_90_085421,prl_113_186102,apl_108_193102} and the enhancement of bond-like contrast in particular \cite{acsnano_10_1201} the flexibility of the CO-terminated tip was assumed and indirectly inferred from the modification of the apparent bond lengths in AFM images of molecules \cite{prl_106_046104,science_336_444}.

These works demonstrate that the exploration of interatomic interactions at bonding distances is most desirable.
With an AFM this range of mutual atom separations can be accessed with picometer precision.
In two recent seminal publications the transition from physisorption to chemisorption was reported in AFM experiments \cite{science_366_235,prl_124_096001}.
By measuring the distance-dependent vertical force between a CO molecule terminating the AFM probe and a single Fe atom adsorbed on a Cu(111) surface two minima separated by a force barrier were observed and assigned to the physisorbed and chemisorbed state of CO and the adsorbed atom \cite{science_366_235}.
It was later shown that such force data may be used to probe the chemical reactivity at the atomic scale \cite{prl_124_096001}.
However, in the light of previous investigations into the flexibility of the CO tip \cite{prl_106_046104,science_336_444} and the reduced stiffness of CO at the tip \cite{science_343_1120} the natural question arises as to the impact of that flexibility on the evolution of the vertical force with the tip--suface distance.
Given the emerging trend of functionalizing scanning probes with even more complex molecules the molecular flexibility at the tip represents an important issue to be explored.

The AFM and scanning tunneling microscope (STM) experiments reported here shed new light on extrema occuring in vertical-force traces.
Using a CO-terminated AFM tip and a Cu(111) surface the measured force-versus-distance evolution is characterized by a dip-hump structure that is similar to the physisorption minimum and the subsequent force barrier reported previously \cite{science_366_235,prl_124_096001}.
The initial CO tilt angle the molecule adopts at the tip apex prior to the tip--surface approach is additionally found to affect the dip-hump structure.
These observations can be described by the gradual bending of the CO molecule in the course of reducing the tip--surface distance.
The underlying simulations rely on the CO deflection in a lateral harmonic, a long-range van der Waals and a short-range Pauli repulsion potential where the latter respects the angular dependence of the orbital overlap.

\begin{figure*}[t]
\includegraphics[width=0.95\linewidth]{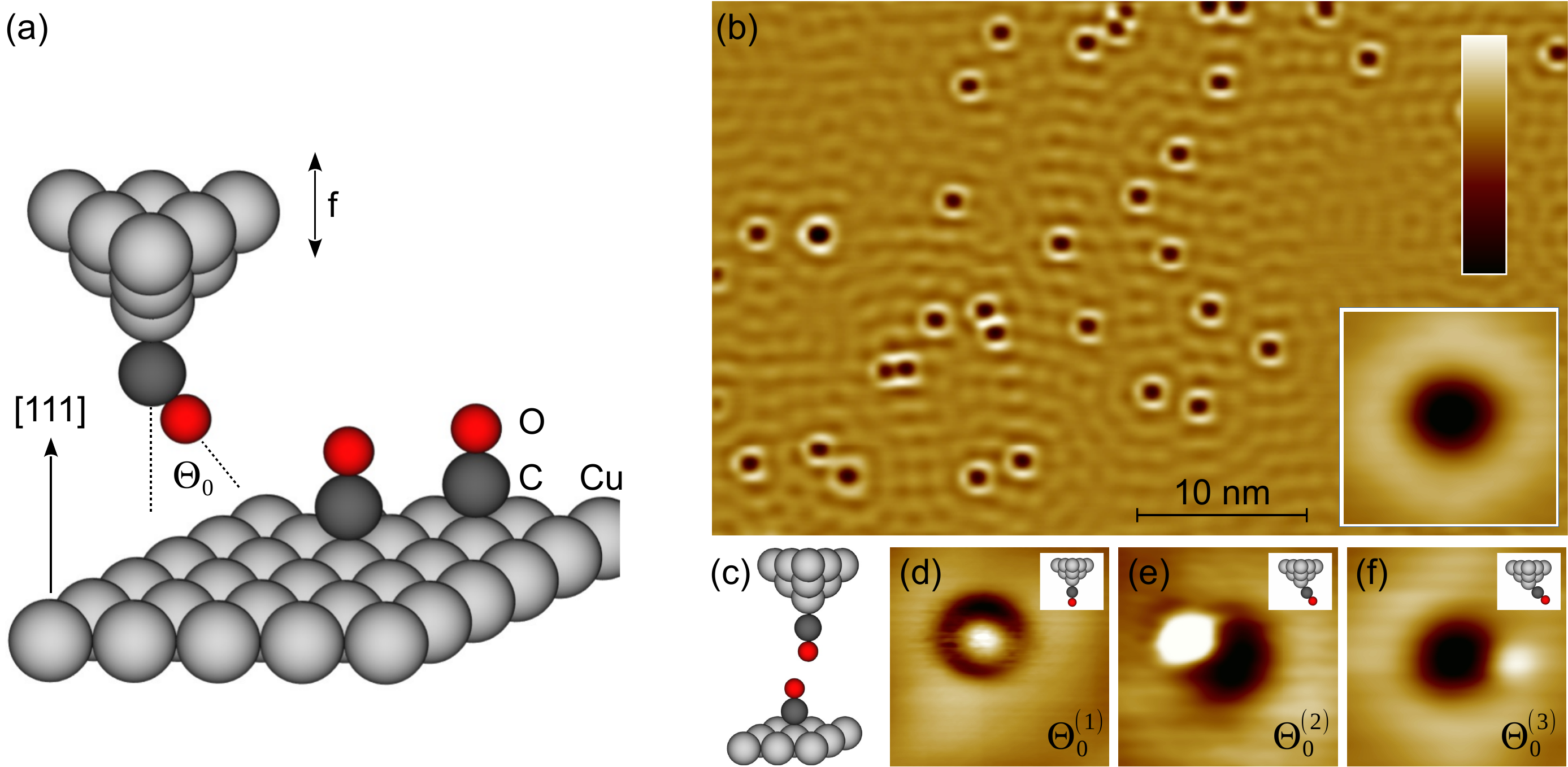}
\caption{\textbf{STM imaging of CO-covered Cu(111)\@.}
(a) Illustration of the experimental setup.
The tip is terminated by a single CO molecule whose axis encloses the initial tilt angle $\Theta_0$ with the $[111]$ surface normal.
In the AFM mode the tip oscillates at frequency $f$.
(b) STM image of the Cu(111) surface with adsorbed CO molecules appearing as circular depressions when recorded with a clean Cu-coated metal tip (bias voltage: $50\,\text{mV}$, tunneling current: $100\,\text{pA}$, size: $50\,\text{nm}\times 30\,\text{nm}$, the color scale ranges from $0\,\text{pm}$ (black) to $150\,\text{pm}$ (white))\@.
The undulation visible across the entire image is due to the electron standing wave pattern induced by the scattering of the Cu(111) Shockley surface state electrons from the CO molecules. 
Inset: Close-up view of a single CO molecule ($2\,\text{nm}\times 2\,\text{nm}$)\@.
(c) Illustration of a tunneling junction comprised of a CO-terminated tip and a CO molecule adsorbed on Cu(111)\@.
(d)--(f) STM images of adsorbed CO molecules recorded with a CO-functionalized tip ($50\,\text{mV}$, $100\,\text{pA}$, $2\,\text{nm}\times 2\,\text{nm}$)\@.
The CO molecule at the tip apex exhibits initial tilt angles $0^\circ\approx\Theta_0^{(1)}<\Theta_0^{(2)}<\Theta_0^{(3)}$\@. 
}
\label{fig1}
\end{figure*}

Figure \ref{fig1}a shows an illustration of the experiment where the tip of the combined AFM-STM is functionalized with a single CO molecule.
The transfer of the molecule to the tip is performed by a standard method \cite{apl_71_213} using CO molecules that had been adsorbed on Cu(111) prior to functionalization (Figure \ref{fig1}b)\@.
Depending on the actual transfer process the C--O axis encloses different initial angles $\Theta_0$ with the surface normal (Figure \ref{fig1}a)\@.
The variety of CO angles may be due to deviations of the actual tip geometry from a perfect pyramidal structure; indeed, tilted adsorption of CO has likewise been reported for differently stepped surfaces \cite{surfsci_415_131}.
The tilted CO configuration at the tip can be imaged by scanning the CO-terminated tip across a single adsorbed CO molecule (Figure \ref{fig1}c--f)\@.
The latter binds with its C atom to an on-top Cu(111) site and is oriented parallel to the surface normal \cite{surfsci_161_349}.
For $\Theta_0\approx 0^\circ$ the STM image reveals a protrusion that is almost concentrically embedded in a depression (Figure \ref{fig1}d)\@.
The more the tilt angle deviates from $0^\circ$ the farther appears the protrusion from the center of the depression (Figure \ref{fig1}e,f)\@.
Depressions and protrusions in STM images of CO acquired with pristine (inset to Figure \ref{fig1}b) and CO-terminated (Figure \ref{fig1}d--f) Cu tips are well understood in terms of the interference of tunneling channels due to the overlap of  \textit{s}-type Cu and \textit{p}-type O orbitals \cite{prb_93_115434,prb_96_085415}.
Previously, however, very often exclusively CO orientations parallel to the surface normal were considered.
Only in scarce cases tilted CO configurations were studied and demonstrated to lead to asymmetric features in force maps recorded around adsorbed molecules \cite{prb_90_085421} and atoms \cite{prb_98_195409,science_366_235}.

\begin{figure}[t]
\includegraphics[width=0.5\linewidth]{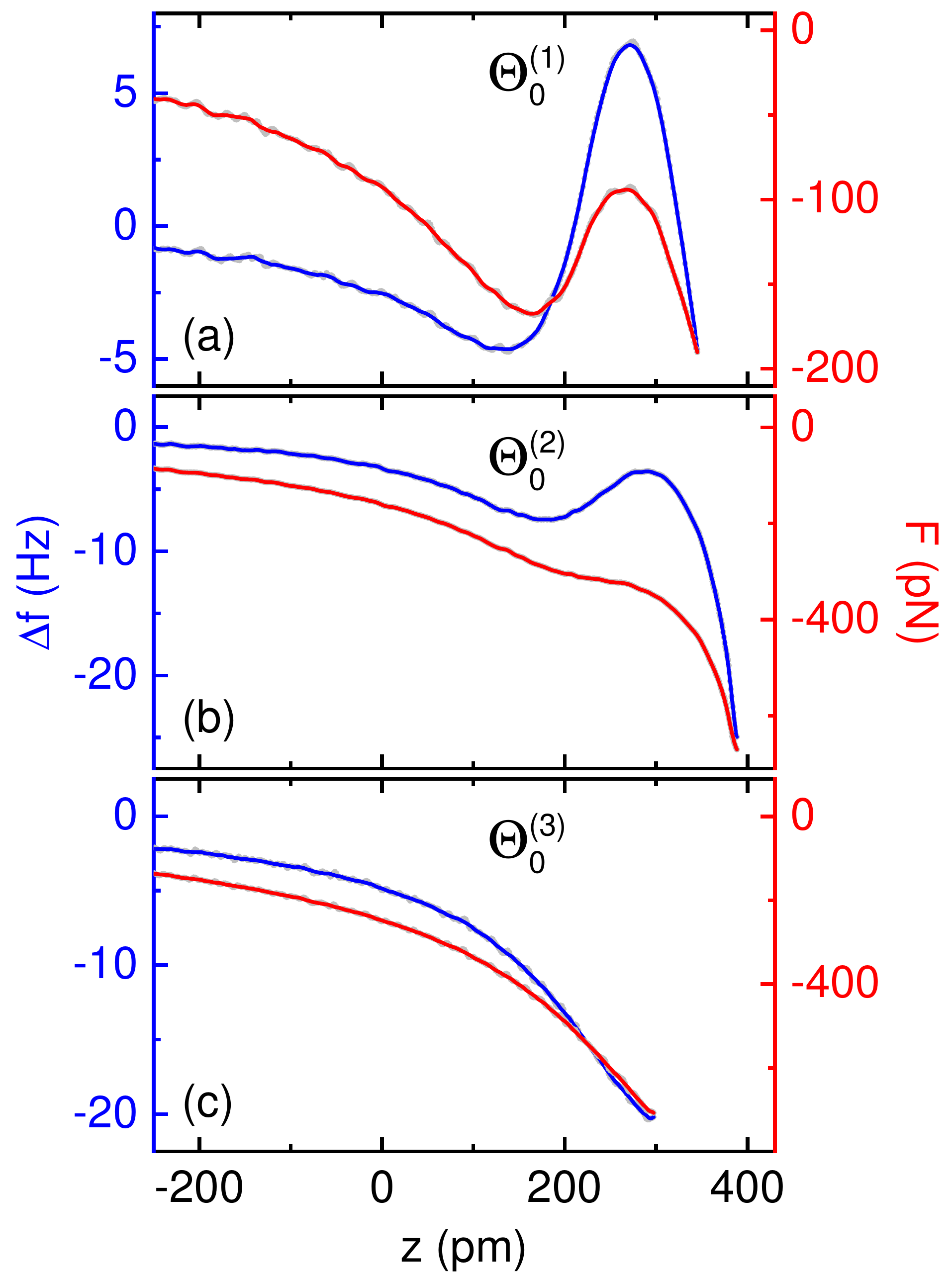}
\caption{\textbf{Distance dependence of CO--surface interaction for different initial CO tilt angles.}
(a)--(c) Frequency shift $\Delta f$ (blue) and vertical force $F$ (red) versus tip displacement $z$ acquired with the CO-terminated tips of Figure \ref{fig1}d--f atop the pristine Cu(111) surface.
Zero displacement is defined by the tip--surface distance prior to disabling the feedback loop at $50\,\text{mV}$ and $100\,\text{pA}$\@.
}
\label{fig2}
\end{figure}

Tips terminated by a single CO molecule were used in the experiments reported here to explore the evolution of the vertical force between the tip and the pristine Cu(111) surface with the tip--surface distance.
To this end mutual separations ranging from far tunneling to chemical-bond distances were covered.
Figure \ref{fig2}a--c shows frequency changes, $\Delta f$, of the oscillating tuning fork and the resulting vertical force, $F$, as a function of the tip displacement, $z$, and for different initial tilt angles $\Theta_0^{(i)}$ ($i=1,2,3$)\@.
Figure \ref{fig2} concentrates on small, intermediate and large initial tilt angles, although a variety of tilt angles was observed.
The Supporting Information present a histogram of the statistical distribution of tilt angles (Figure S1)\@.
Zero displacement defines the tip--surface distance prior to disabling the STM feedback loop at identical bias voltage ($50\,\text{mV}$) and tunneling current ($100\,\text{pA}$) for all tips.
The tip was then retracted to a common initial height ($z<0\,\text{pm}$)\@.
Except for tips with large initial CO tilt angles (Figure \ref{fig2}c) $\Delta f$ and $F$ evolve in a significantly different manner than expected from the Lennard-Jones-type behavior.
For $\Theta_0^{(1)}\approx 0^\circ$ (Figure \ref{fig2}a) the deviations are most pronounced in that $\Delta f(z)$ and $F(z)$ exhibit a characteristic dip-hump structure where a local minimum is followed by a local maximum and a subsequent steep decrease with the reduction of the tip--surface separation.
A larger tilt angle, $\Theta_0^{(2)}>\Theta_0^{(1)}$, yields a dip-hump evolution of $\Delta f(z)$ and $F(z)$, too, albeit in a less pronounced manner (Figure \ref{fig2}b)\@.
Even larger tilt angles, $\Theta_0^{(3)}>\Theta_0^{(2)}$, cause the suppression of the peculiar structure in both $\Delta f(z)$ and $F(z)$ (Figure \ref{fig2}c)\@.
For all initial tilt angles, further approach of the tip to the surface led to abrupt changes in $\Delta f$ and material transfer from the tip to the surface signaling a jump to contact \cite{prl_94_126102,jpcm_20_223001}.

The comparison of $F(z)$ traces for $\Theta_0^{(1)}$ (Figure \ref{fig2}a) and $\Theta_0^{(2)}$ (Figure \ref{fig2}b) shows that the $\Theta_0^{(1)}$-tip exhibits a lower attractive van der Waals attractive force than the $\Theta_0^{(2)}$-tip, e.\,g., $\approx -100\,\text{pN}$ ($\Theta_0^{(1)}$) and $\approx -200\,\text{pN}$ ($\Theta_0^{(2)}$) at $z=0\,\text{pm}$.
Therefore, the magnitude of the long-range attraction, rather than $\Theta_0$, may in principle influence the actual dip-hump variation of $F(z)$\@.
The Supporting Information (Figure S2), however, exclude this scenario by presenting pronounced dip-hump structures as well for tips with elevated van der Waals forces.
Moreover, the presence of the CO molecule within the junction over the whole range of tip displacements was evidenced by inelastic electron tunneling spectroscopy revealing vibrational modes associated to the CO molecule (Supporting Information, Figure S3)\@.

Because of missing atomic resolution of the Cu(111) lattice a statistical ensemble of substrate sites was presumably probed in the contact experiments.  
The acquired $\Delta f(z)$ data for a given CO-terminated tip were very similar and, therefore, support the idea of a weak impact of the contacted surface site on the induced CO deflection.
Indeed, differences in $F(z)$ data are on the order of $10\,\text{pN}$ (Supporting Information, Figure S4)\@.
Another example of weak site-specific force evolutions was reported for CO-terminated tips above Cu$_{2}$N islands on Cu(100) \cite{prl_112_166102}.

The presented dip-hump evolution of $\Delta f(z)$ and $F(z)$ as well as its dependence on the initial CO tilt angle at the tip bear resemblance to previous results reported for single Fe atoms and Fe clusters adsorbed on Cu(111) \cite{science_366_235,prl_124_096001}.
In these works \cite{science_366_235,prl_124_096001} the occurrence of a shallow and a deep minimum of $F(z)$ separated by a force barrier ($>0\,\text{pN}$) was rationalized in terms of the physisorbed and chemisorbed state of CO and the adsorbates \cite{science_366_235}.
Moreover, it was found that different CO-terminated tips yielded differently pronounced physisorption minima and force barriers without further describing the actual differences of the tips.
In the following, without resorting to the presence of physisorption and chemisorption, a model is introduced that describes the characteristic dip-hump evolution of $F(z)$ reported here and explains its dependence on the initial CO tilt angle.

\begin{figure}
\includegraphics[width=0.2\linewidth]{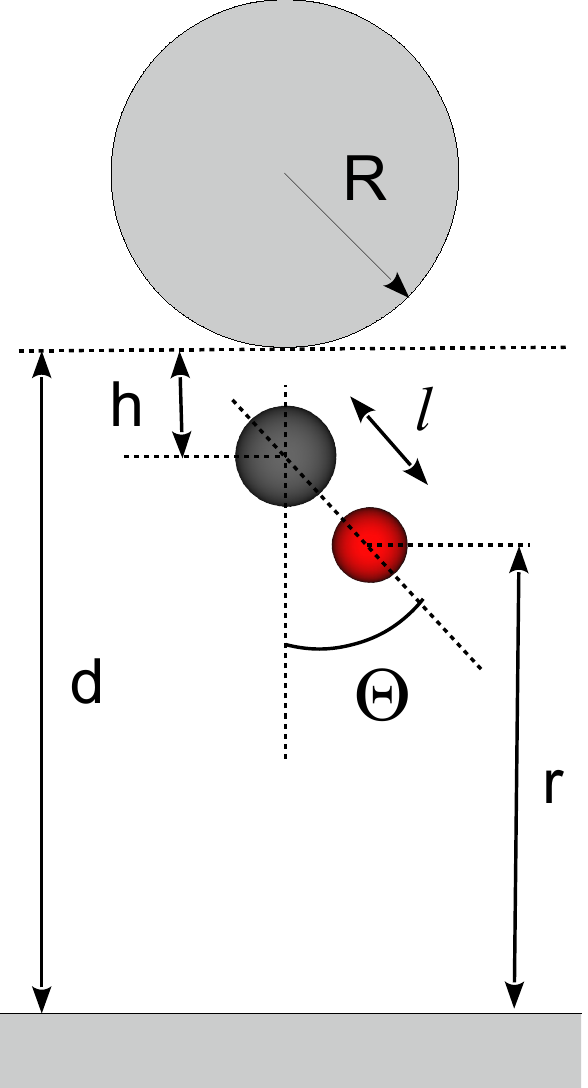}
\caption{\textbf{Illustration of the model geometry.}
The tip is described as a metal sphere with radius $R$, a distance $d$ from the semiinfinite flat substrate.
The CO molecule is attached to the tip via the C atom at a distance $h$ and exhibits an equilibrium bond length $\ell$.
The O--surface separation is $r$.
The tilt angle $\Theta$ and the distance $r$ are functions of $d$.
}
\label{fig3}
\end{figure}

A sketch of the model setup is presented in Figure \ref{fig3}\@.
The CO molecule may rotate in a harmonic potential
\begin{equation}
V_{\text{rot}}=\frac{1}{2}kl^2\sin^2(\Theta-\Theta_0)
\label{eq:vrot}
\end{equation}
with $k$ the lateral spring constant, $l$ the CO equilibrium bond length and $\Theta=\Theta(d)$ the CO tilt angle that depends on the distance $d$ between a metallic sphere of radius $R$ representing the tip and the metal surface.
This potential describes the frustrated translation of CO on Cu(111) \cite{science_336_444}.
Additionally, the van der Waals potential
\begin{equation}
V_{\text{vdW}}=-H\frac{R}{6d}
\label{eq:vvdw}
\end{equation}
($H$: Hamaker constant) was taken into account for modeling the long-range and slowly varying interaction between the macroscopic tip and the surface.
The interaction of the CO molecule and the surface was simulated by the Lennard-Jones potential that simplifies the O and Cu atoms as point-like objects separated by a distance $r=r(d)$, i.\,e.,
\begin{equation}
V_{\text{LJ}}=\varepsilon\left[\left(\frac{r_0}{r}\right)^{12}-2\left(\frac{r_0}{r}\right)^6\right]
\label{eq:vlj}
\end{equation}
($\varepsilon$: depth of the Lennard-Jones potential well, $r_0$: equilibrium Cu--O bond distance)\@.
The long-range electrostatic potential was neglected due to the order-of-magnitude lower associated force and its slow variation with $d$ compared to the van der Waals force in the relevant distance range close to maximum attraction.

For each distance $d$ in the simulations the minimum of the total potential, $V=V_{\text{rot}}+V_{\text{vdW}}+V_{\text{LJ}}$, with respect to variations in $\Theta$, $\left.\partial V/\partial\Theta\right|_{\Theta=\Theta^\ast}=0$, was determined and the resulting vertical force extracted, $F=-\partial V^\ast/\partial d$ with $V^\ast=V(\Theta^\ast)$\@.
This procedure was repeated for different initial tilt angles $\Theta_0$\@.
In order to facilitate the comparison with experimental, $d=-z$, $H=0.5\,\text{eV}$ \cite{rsi_64_1868}, $R=10\,\text{nm}$ \cite{rsi_62_2167}, $\varepsilon=0.1\,\text{eV}$ and $r_0=0.35\,\text{nm}$ \cite{prb_90_085421} were set.
In addition, $k=0.85\,\text{N}/\text{m}$ was adopted from previous density functional calculations \cite{natnanotechnol_13_371}.
Distances $h=0.191\,\text{nm}$ (Ref.\,\cite{jpcm_16_1141}) between C and the tip apex (Figure \ref{fig2}a) as well as $\ell=0.11\,\text{nm}$ (Ref.\,\cite{pr_78_140}) were left invariant in the calculations.

\begin{figure}
\includegraphics[width=0.45\linewidth]{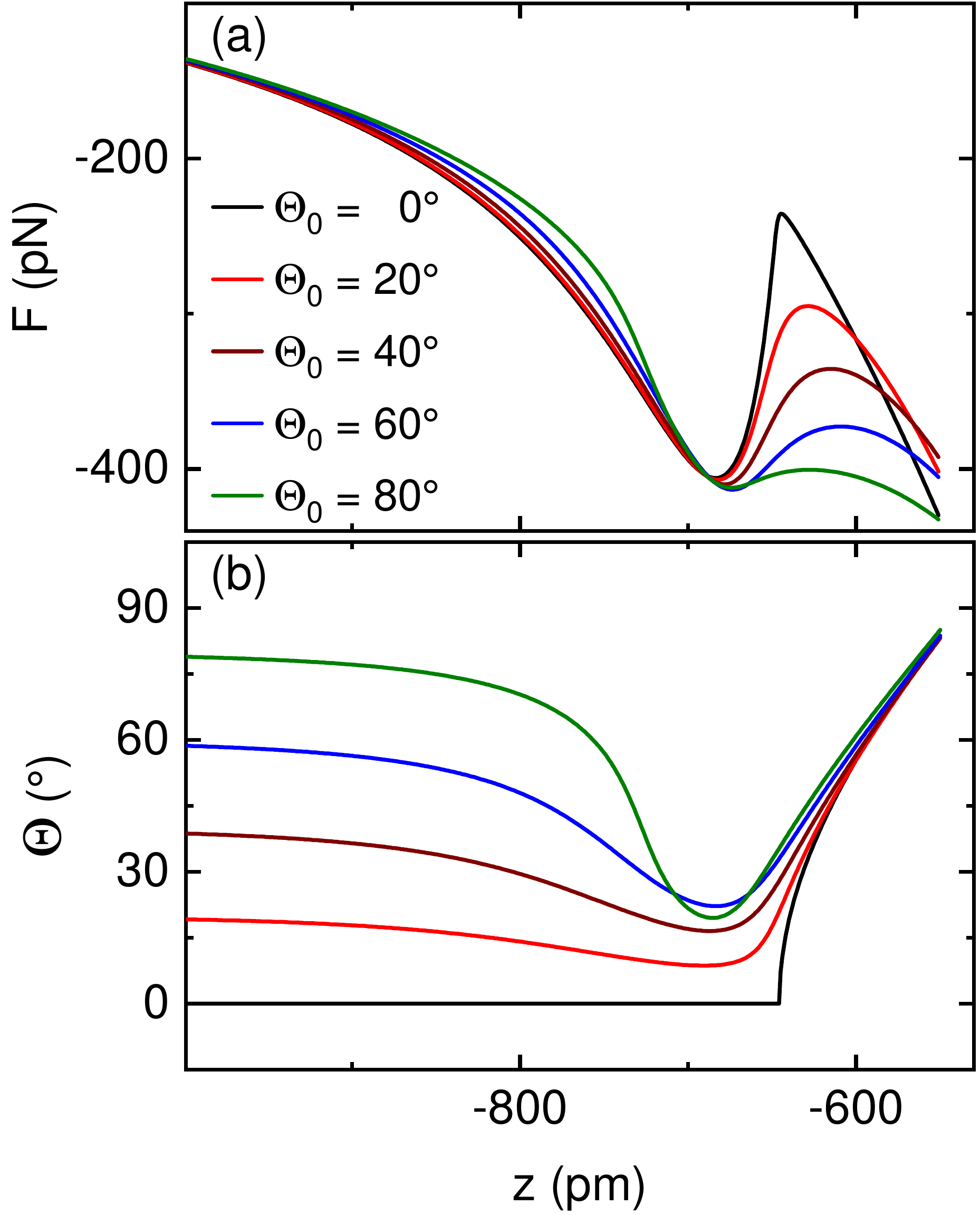}
\caption{\textbf{Simulation of the CO rotational motion.}
(a) Vertical force $F$ derived from potentials in eqs \ref{eq:vrot}--\ref{eq:vlj} for the indicated initial CO tilt angles $\Theta_0$\@.
(b) Variation of the tilt angle $\Theta$ with distance $d=-z$ for different $\Theta_0$\@.
}
\label{fig4}
\end{figure}

Figure \ref{fig4}a shows calculated vertical forces for different $\Theta_0$\@.
The dip-hump structure observed in the experimental $F(z)$ data is qualitatively reproduced.
The local minimum of $F$ is due to contact of the O atom of CO with the surface and is mainly caused by $V_{\text{LJ}}$\@.
This minimum nearly coincides with the tip--surface distance where the tilt angle $\Theta$ changes from bending towards the surface normal -- mediated by a torque due to the attractive components of the interaction potential -- to tilting away from it -- mediated by a torque due to the Pauli repulsion (Figure \ref{fig4}b)\@.
The reaching of a local maximum followed by a renewed decrease of $F(z)$ is due to the continuation of attractive van der Waals forces in the course of further approaching the tip.
Without the flexibility of the single-molecule junction entailed by the CO rotational motion the attractive van der Waals forces would partially be compensated by the short-range Pauli repulsion. 

The experimental data (Figure \ref{fig2}) unveil that the local force maximum becomes progressively attenuated with increasing $\Theta_0$\@.
This decrease of the maximum is reproduced in the simulation, albeit to a less pronounced extent.
The calculations therefore overestimate the repulsive forces present in the experiments.
In fact, the repulsive term of the Lennard-Jones potential depends on the distance between the O atom and the surface, which is strictly valid only for a spherical electron distribution around the O atom.
However, the actual electron distribution deviates from a sphere due to accumulated electron density along the CO axis \cite{natnanotechnol_13_371}.
Therefore, it is reasonable to assume that the Pauli repulsion term of the Lennard-Jones potential depends on the orientation of CO with respect to the surface.

\begin{figure}
\includegraphics[width=0.45\linewidth]{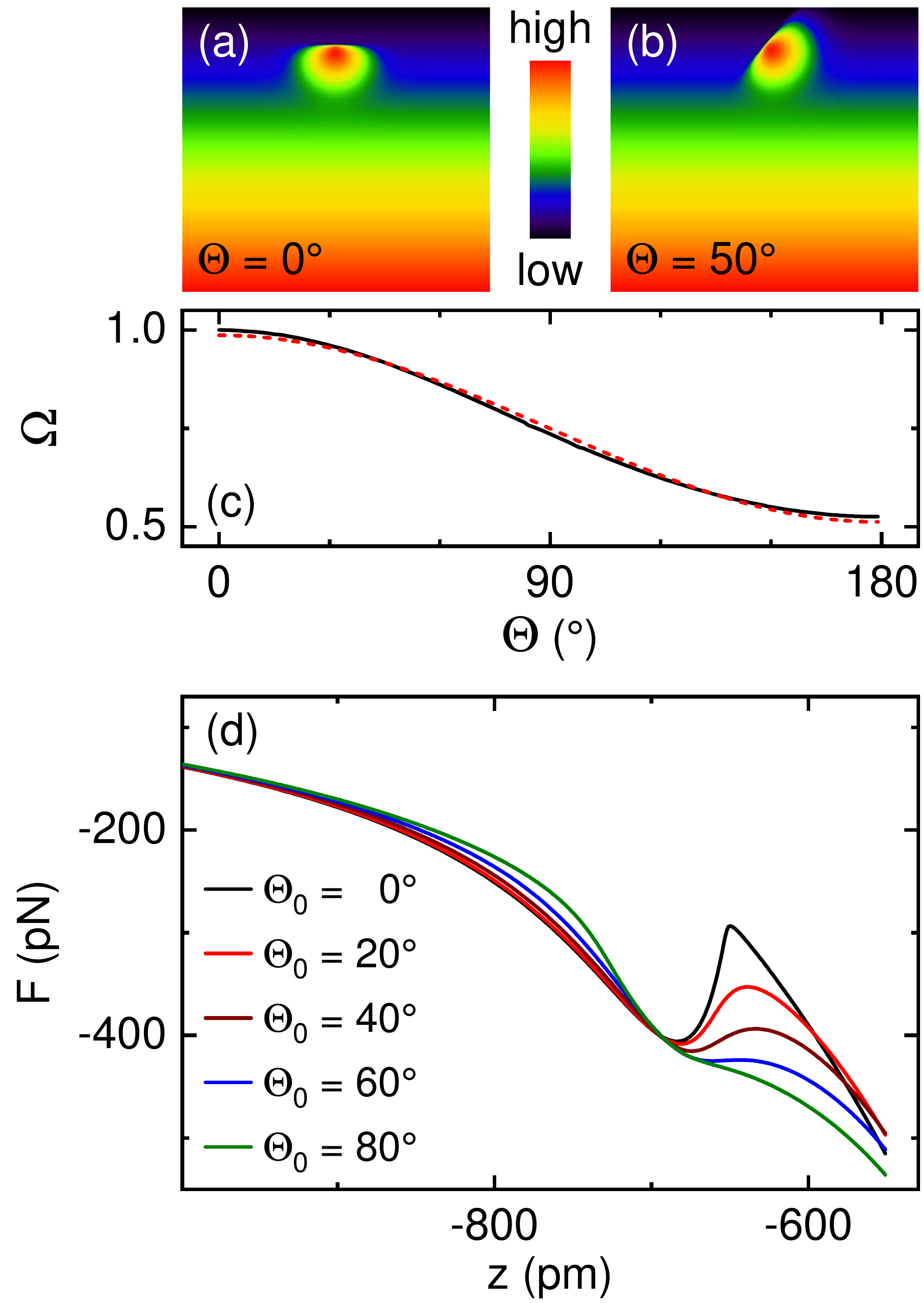}
\caption{\textbf{CO tilt angle dependence of the orbital overlap.}
(a), (b) Illustration of the overlap of electron densities due to a single $p_z$-orbital $\psi_{p_z}$ and a surface $s$-wave function $\psi_s$ for $\Theta=0^\circ$ (a) and $\Theta=50^\circ$ (b)\@.
(c) Overlap $\Omega=\int\psi_{p_z}\psi_s\,\text{d}V$ as a function of $\Theta$\@.
The dashed line depicts the graph of $g(\Theta)=1+a(\cos\Theta-1)$ with $a=0.25$\@.
(d) Vertical force $F$ derived from the sum of potentials in eqs \ref{eq:vrot},\ref{eq:vvdw},\ref{eq:vljmod} where the $\Theta$ dependence of the Pauli repulsion term in the Lennard-Jones potential is considered, $g(\Theta)$ with $a=0.4$\@.
}
\label{fig5}
\end{figure}

In order to empirically determine the angle dependence of the Pauli repulsion term the overlap integral $\Omega=\int\psi_{p_z}\psi_s\,\text{d}V$ of a $p_z$ molecular orbital with an $s$-type wave function of the surface, $\psi_s\propto\exp(-\kappa z)$ with a spatial decay constant $\kappa$, was calculated at a given tip--surface distance and for different rotation angles.
Electron densities are shown in Figure \ref{fig5}a,b.
The overlap integral was calculated considering only the lower lobe of the $p_z$ molecular orbital in order to mimic the molecular orbital of the CO molecule pointing toward the surface.
The variation of the overlap with the CO tilt angle, $\Omega=\Omega(\Theta)$, is sinusoidal (Figure \ref{fig5}c)\@.
Owing to these results an angle-dependent prefactor of the Pauli repulsion term was introduced, which gives rise to a modified Lennard-Jones potential of the form
\begin{equation}
\tilde{V}_{\text{LJ}}=\tilde{V}_{\text{LJ}}(r;\Theta)=\varepsilon\left[g(\Theta)\left(\frac{r_0}{r}\right)^{12}-2\left(\frac{r_0}{r}\right)^6\right]
\label{eq:vljmod}
\end{equation}
with $g(\Theta)=1+a(\cos\Theta-1)$\@.
The simulated force data are depicted in Figure \ref{fig5}d for different angles and $a = 0.4$.
The attenuation of the Pauli repulsion term due to $g(\Theta)$ leads to an improved description of the experimental data.

Summarizing the results of the simulations, the occurrence of extrema in the vertical force can be understood in the following way.  
The local minimum is due to the contact of the O atom of the CO-terminated tip with the Cu(111) surface.  
The subsequent increase of $F$ is caused by the Pauli repulsion.  
The Supporting Information (Figure S5) demonstrate that the maximum of $F$ appears because the CO molecule bends away from the surface normal and, thereby, allows the macroscopic tip to further approach the surface without significantly reducing the O--Cu separation.
Therefore, the van der Waals attraction concomitantly overcompensates the Pauli repulsion leading to the observed steep decrease of $F$ after reaching the maximum.
The Supporting Information likewise show that for large $\Theta_0$ substantial energy is gained due to the bending of the molecule back to its equilibrium orientation for tip approach beyond the point of maximum attraction, while for small $\Theta_0$ the CO molecule tilts away from $\Theta_0$\@.
Therefore, the Pauli repulsion is efficiently reduced for large $\Theta_0$ owing to comparably small changes of the O--Cu distance.

The force evolution $F(z)$ for a CO-terminated tip approaching a single adsorbed Cu atom is markedly different (Supporting Information, Figure S6) because it exhibits the expected Lennard-Jones behavior with a minimum reflecting the Cu--O bond formation and a subsequent steep monotonous increase due to Pauli repulsion, which is in agreement with a previous report \cite{science_366_235}.
The bending of CO as observed in the case of the atomically flat Cu(111) surface, therefore, is efficiently suppressed due to bond formation to the adsorbed Cu atom.
The protruding Cu atom on the surface offers a directional bond to the O atom, resulting most likely in a $\text{Cu}\,d_{z^2}$--$\text{O}\,p_z$ hybrid orbital \cite{science_366_235}.
In contrast, the bare surface gives rise to a spatially uniform charge density that impedes the anchoring of the CO molecule at a specific surface site, thereby enabling its deflection under the influence of the approaching tip.

In conclusion, a dip-hump evolution of the vertical force between a CO-terminated probe and a surface in the range of chemical-bond distances reflects the gradual deflection of the CO molecule induced by the joint action of torques due to long-range van der Waals attraction and short-range Pauli repulsion.
The CO bending is described in a lateral harmonic potential while the attenuation of the dip-hump structure is captured by considering the CO orientation in the Lennard-Jones potential.
The molecular functionalization of scanning probes is an emerging tool to improve spatial resolution in imaging \cite{science_325_1110}, to use single-molecule vibrational quanta for the real-space study of chemical bonds \cite{science_344_885_a}, to explore atomic-scale reactivity \cite{science_366_235_a,prl_124_096001} or to access single-atom spin excitations \cite{science_364_670,science_366_623}.
The findings reported in the present work demonstrate that the molecular probe is subject to bending.
The resulting change in the molecular orientation is likely to affect the probed properties.

\textbf{Experimental method.} Experiments were performed with an apparatus combining an AFM and an STM operated in ultrahigh vacuum ($10^{-9}\,\text{Pa}$) and at low temperature ($5\,\text{K}$).
The vertical force between tip and surface was extracted from non-contact frequency modulation measurements.
PtIr tips were attached to the free prong of a quartz tuning fork \cite{apl_73_3956} oscillating with a resonance frequency of $29.9\,\text{kHz}$, a quality factor of $45000$ and an amplitude of $50\,\text{pm}$.
Prior to the experiments the AFM tip had been prepared by focused ion beam milling to ensure a well defined shape of the apex.
In situ, the PtIr tip was prepared by field emission on and indentation into the Cu surface, which presumably led to a coating of the tip apex with several atomic layers of the substrate material.
The Cu(111) surface was prepared by Ar$^+$ bombardment and annealing.
CO molecules were deposited on the cold surface by opening the radiation shields for $30\,\text{s}$ while the vacuum vessel was backfilled with gaseous CO (purity: $99.995\,\%$) at a partial pressure of $10^{-7}\,\text{Pa}$.
The vertical force was calculated from the measured resonance frequency variation, $\Delta f$, using different deconvolution methods \cite{apl_78_123,apl_84_1801} that led to virtually identical results.
STM images of the sample surfaces were acquired in the constant-current mode with the bias voltage applied to the sample.
Topographic data were processed using WSxM \cite{rsi_78_013705}.

\begin{acknowledgement}
Discussions with Magnus Paulsson (Linnaeus University, Sweden), Manuel J.\ Kolb (Massachusetts Institute of Technology, USA) and financial support by the Deutsche Forschungsgemeinschaft through grant no.\,KR 2912/12-1 are acknowledged.
\end{acknowledgement}

\begin{suppinfo}
The Supporting Information are available free of charge on the ACS Publications website at DOI: [hyperlink DOI] \\
Statistical distribution of initial CO tilt angles; CO deflection and van der Waals attraction; Inelastic electron tunneling spectroscopy; Dependence of force evolution on Cu(111) lattice sites; Origin of force extrema; Comparison between Cu(111) and adsorbed Cu atom.
\end{suppinfo}

\providecommand{\latin}[1]{#1}
\makeatletter
\providecommand{\doi}
  {\begingroup\let\do\@makeother\dospecials
  \catcode`\{=1 \catcode`\}=2 \doi@aux}
\providecommand{\doi@aux}[1]{\endgroup\texttt{#1}}
\makeatother
\providecommand*\mcitethebibliography{\thebibliography}
\csname @ifundefined\endcsname{endmcitethebibliography}
  {\let\endmcitethebibliography\endthebibliography}{}

\renewcommand{\thefigure}{S\arabic{figure}}
\renewcommand{\thetable}{S\arabic{table}}
\renewcommand{\thepage}{S\arabic{page}}
\renewcommand{\theequation}{S\arabic{equation}}

\newpage

\section{Supporting Information: Atomic Force Extrema Induced by the Bending of a CO-Functionalized Probe}

\section{Statistical distribution of initial CO tilt angles}

\begin{figure}
\centering
\includegraphics[width=0.6\linewidth]{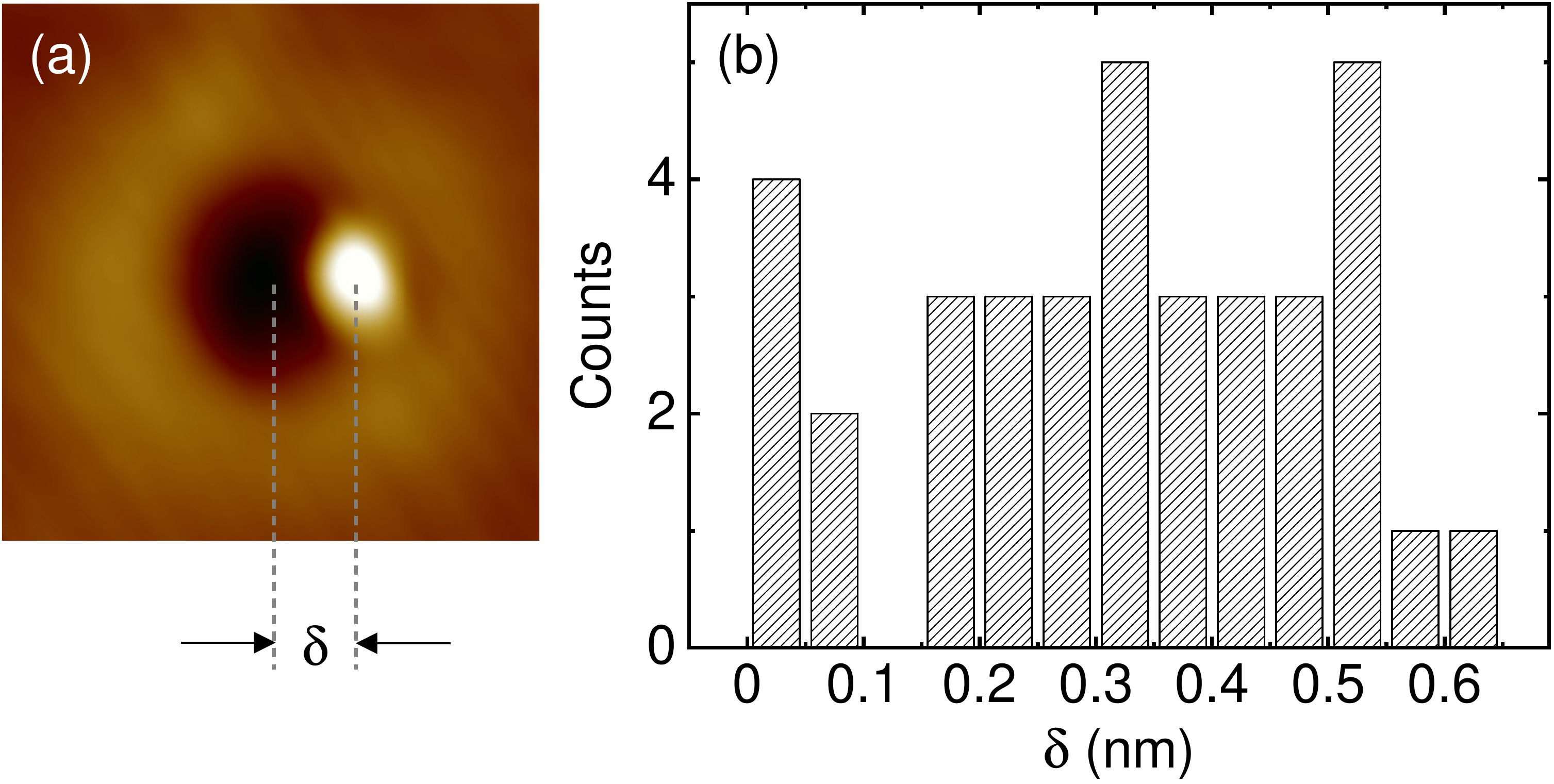}
\caption{
(a) STM image of a single CO molecule adsorbed on Cu(111) recorded with a CO-terminated tip ($50\,\text{mV}$, $100\,\text{pA}$, $2\,\text{nm}\times 2\,\text{nm}$)\@.
Distance $\delta$ between the center of the protrusion and the center of the depression is indicated\@.
Histogram of $\delta$ for a total of 36 tips.
Bins of width $0.5\,\text{nm}$ are used.
}
\label{figS1}
\end{figure}

In total, 36 CO-terminated tips were fabricated in the experiments.
For each tip a CO molecule adsorbed on Cu(111) was imaged (Figure \ref{figS1}a)\@.
Due to the unknown tip--surface distance the CO tilt angle cannot be extracted.
However, the distance $\delta$ between the center of the protrusion and the center of the depression (Figure \ref{figS1}a) represents a measure of the tilt angle.
The histogram therefore shows the statistical distribution of $\delta$, rather than of $\Theta_0$\@.
It does not reflect a preferential $\delta$.
Remarkably, $\delta$ exceeds the geometric distance between C and O (denoted $l$ in the article)\@.
Most likely, this observation is due to the extended molecular orbitals involved in the tunneling process.

\section{CO deflection and van der Waals attraction} 

\begin{figure}
\centering
\includegraphics[width=0.5\linewidth]{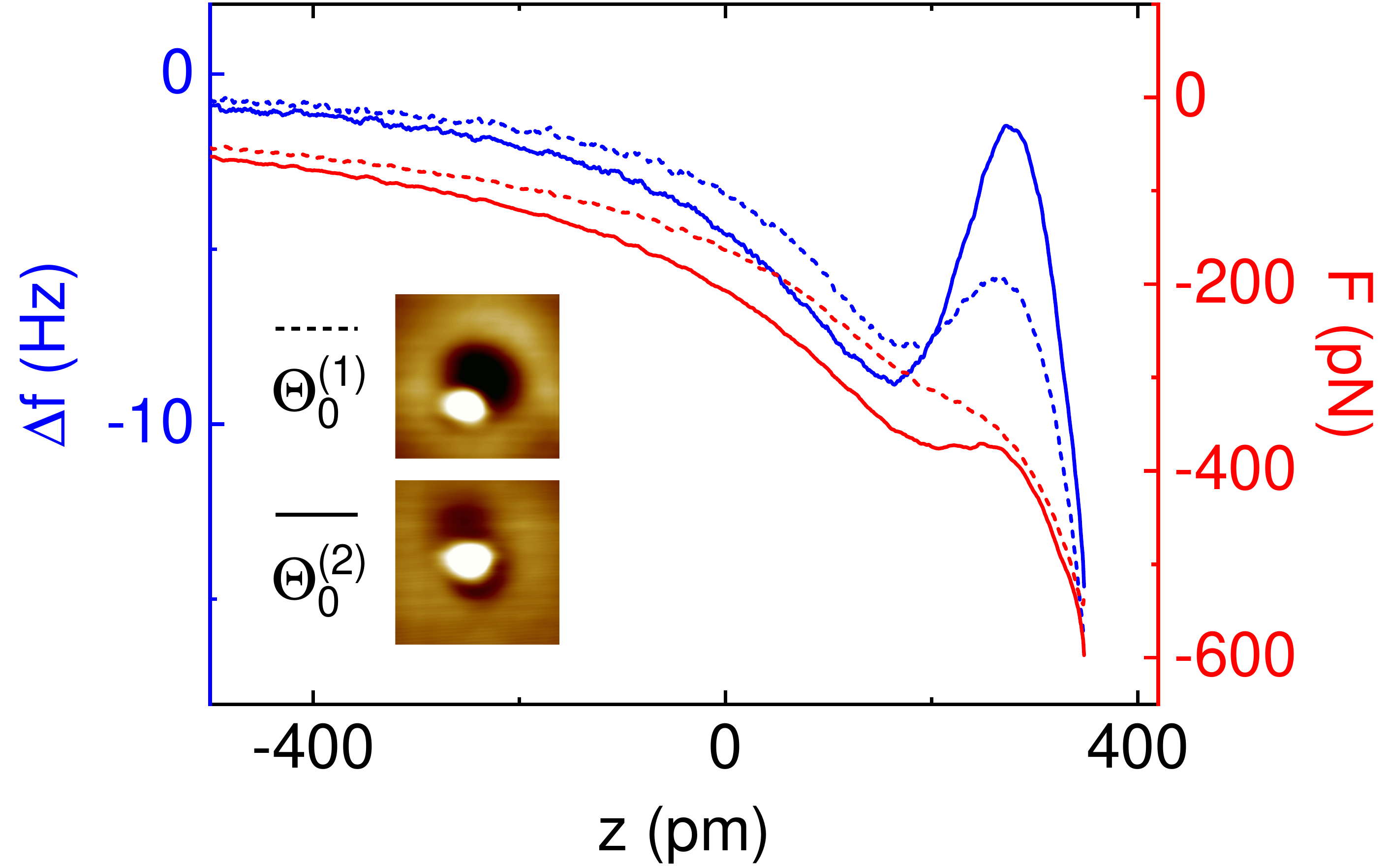}
\caption{
Frequency shift $\Delta f$ (blue) and force $F$ (red) versus tip displacement $z$ for two CO-terminated tips with initial tilt angles $\Theta_0^{(1)}$ (dashed lines) and $\Theta_0^{(2)}$ (solid lines), $\Theta_0^{(1)}>\Theta_0^{(2)}$, recorded above the bare Cu(111) surface.
Zero displacement ($z = 0\,\text{pm}$) corresponds to the tip--surface distance prior to disabling the feedback loop ($50\,\text{mV}$, $100\,\text{pA}$)\@.
Insets: STM images of a single CO molecule adsorbed on Cu(111) recorded with the CO-terminated tips used for acquiring the $\Delta f$, $F$ data ($50\,\text{mV}$, $100\,\text{pA}$, $2\,\text{nm}\times 2\,\text{nm}$)\@.
}
\label{figS2}
\end{figure}

The characteristic dip-hump structure of $\Delta f(z)$ and $F(z)$ is related to the CO initial tilt angle, $\Theta_0$\@.
Figure 2 of the article shows that smaller initial angles $\Theta_0$ yield more pronounced dip-hump variations.
The data presented in Figure 2a--c were acquired with different tips, which can be inferred from the absolute values of $\Delta f$ and $F$\@.
In particular, the attractive van der Waals interaction for the tip used in Figure 2b is stronger than for the tip leading to the data in Figure 2a.
Therefore, one may argue that the attractive van der Waals force determines the extent of the dip-hump stucture, rather than the initial tilt angle.
Figure \ref{figS2} refutes this assumption.
It shows $\Delta f$ and $F$ for CO-terminated tips with initial tilt angles $\Theta_0^{(1)}$ (dashed lines) and $\Theta_0^{(2)}$ (solid lines)\@.
The STM images (insets to Figure \ref{figS2}) reveal that $\Theta_0^{(1)}>\Theta_0^{(2)}$ while $\Delta f$ and $F$ for the CO-terminated tip with $\Theta_0^{(1)}$ exceed the values for the other tip.
Consequently, Figure \ref{figS2} represents the opposite situation to Figure 2a,b, i.\,e., a stronger attractive van der Waals force together with a pronounced dip-hump structure.
Therefore, the tilt angle $\Theta_0$ rather than the magnitude of the van der Waals attraction determines the characteristic evolution of $\Delta f$ and $F$ with $z$.

\section{Inelastic electron tunneling spectroscopy}

\begin{figure}
\centering
\includegraphics[width=0.6\linewidth]{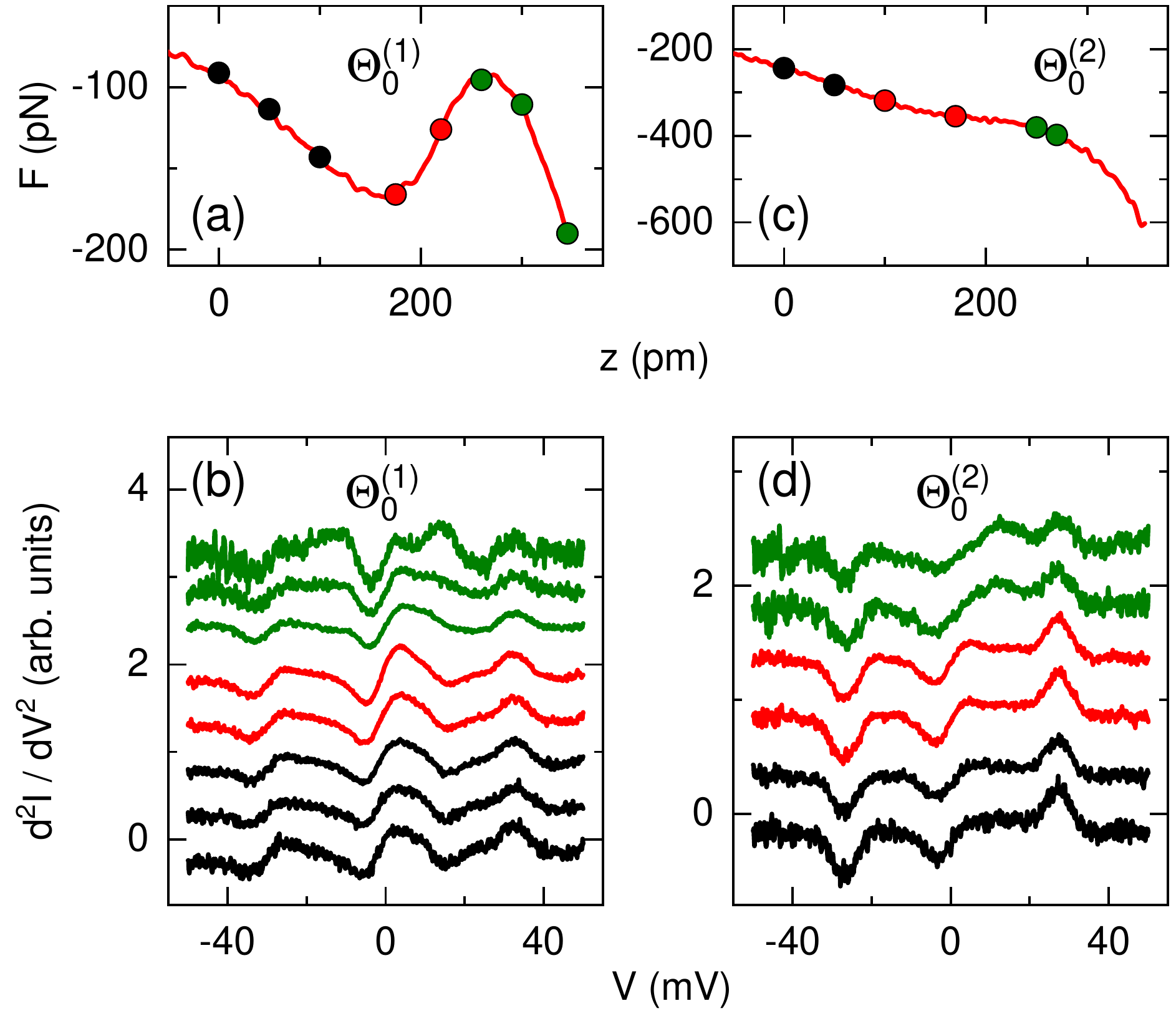}
\caption{
(a)  Force $F$ between a CO-terminated tip with initial tilt angle $\Theta_0^{(1)}$ and clean Cu(111) as a function of the displacement $z$.
(b) Inelastic electron tunneling spectra acquired at the tip displacements indicated by colored dots in (a)\@. 
The data are divided by the current at $50\,\text{mV}$ and are vertically offset for clarity.
(c),(d) Same as (a),(b) for $\Theta_0^{(2)}>\Theta_0^{(1)}$\@.
}
\label{figS3}
\end{figure}

The CO-terminated tip was characterized by inelastic electron tunneling spectroscopy on clean Cu(111) in tunneling and contact distance ranges.
Figure \ref{figS3} presents the results for tips with initial CO tilt angles $\Theta_0^{(1)}$ (Figure \ref{figS3}a,b) and $\Theta_0^{(2)}$ (Figure \ref{figS3}c,d) with $0^\circ\approx\Theta_0^{(1)}<\Theta_0^{(2)}$\@.
In the tunneling distance range the $\text{d}^2I/\text{d}V^2$ spectra reveal dip-peak pairs at $\approx\pm 3\,\text{mV}$ and $\approx\pm 33\,\text{mV}$ for CO-terminated tips with $\Theta_0^{(1)}$\@.
These features are assigned to the frustrated translation and rotation of CO, respectively \cite{prl_87_196102,science_344_885,prb_93_165415,prl_116_166101,prl_118_036801,prl_119_166001}.
The energies of these vibrational modes do not change for the wide tip--surface distances probed, which proves that the CO molecule remains in the junction.

The CO-terminated tip with tilt angle $\Theta_0^{(2)}$ exhibits a slightly lower energy for the frustrated rotation ($\approx 28\,\text{meV}$) than the tip with $\Theta_0^{(1)}$, while the frustrated translation energy is essentially independent of the initial tilt angle.
According to previous calculations the frustrated rotation of CO at the tip apex is characterized by larger displacements of the C atom compared to the O atom \cite{prl_119_166001}.
Since the CO molecule bonds with its C atom to the tip apex the large C displacement renders the frustrated rotation sensitive to the atomic-scale geometry of the tip apex.
For different tips it is therefore reasonable to assume that the frustrated rotation is affected in its energy. 

\section{Dependence of force evolution on Cu(111) lattice sites} 

\begin{figure}
\centering
\includegraphics[width=0.4\linewidth]{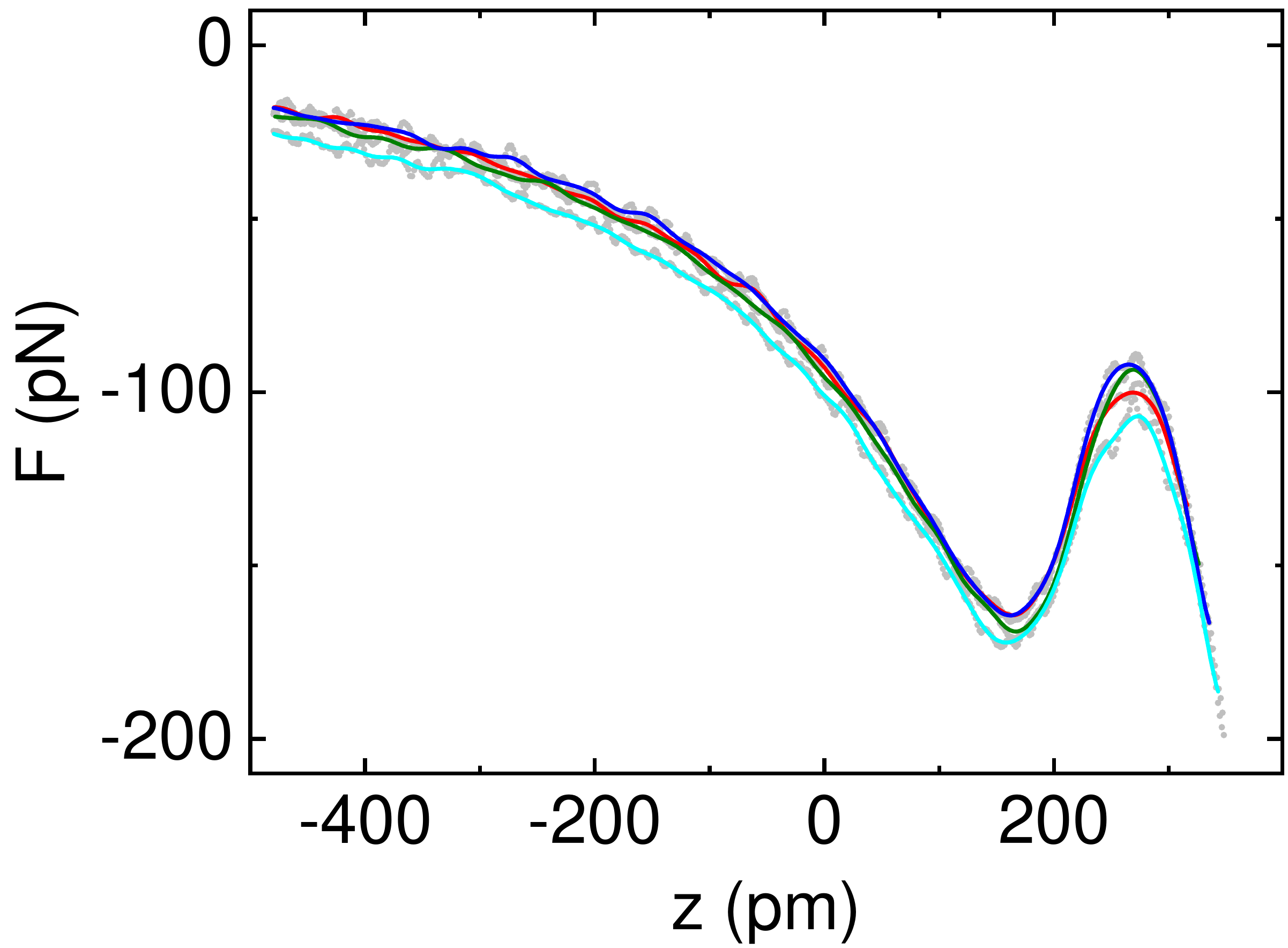}
\caption{
Force $F$ versus tip displacement $z$ for a CO-terminated tip recorded above various positions of the bare Cu(111) surface.
Zero displacement ($z = 0\,\text{pm}$) corresponds to the tip--surface distance prior to disabling the feedback loop ($50\,\text{mV}$, $100\,\text{pA}$)\@.
}
\label{figS4}
\end{figure}

Figure \ref{figS4} shows $F(z)$ data recorded with a single CO-terminated tip above various positions of the bare Cu(111) surface.
Care has been taken to ensure a common initial position of the tip prior to disabling the feedback control.
Due to the lack of atomic resolution of the substrate surface lattice the actual approach site is unknown.
The deviations of the data sets may be attributed to different interactions between the CO-terminated tip and the approached surface sites.
The differences are on the order of $10 \,\text{pN}$ and reflect a weak influence of the specific surface site on the $F(z)$ evolution. 

\section{Origin of force extrema}

\begin{figure}[t]
\centering
\includegraphics[width=0.5\linewidth]{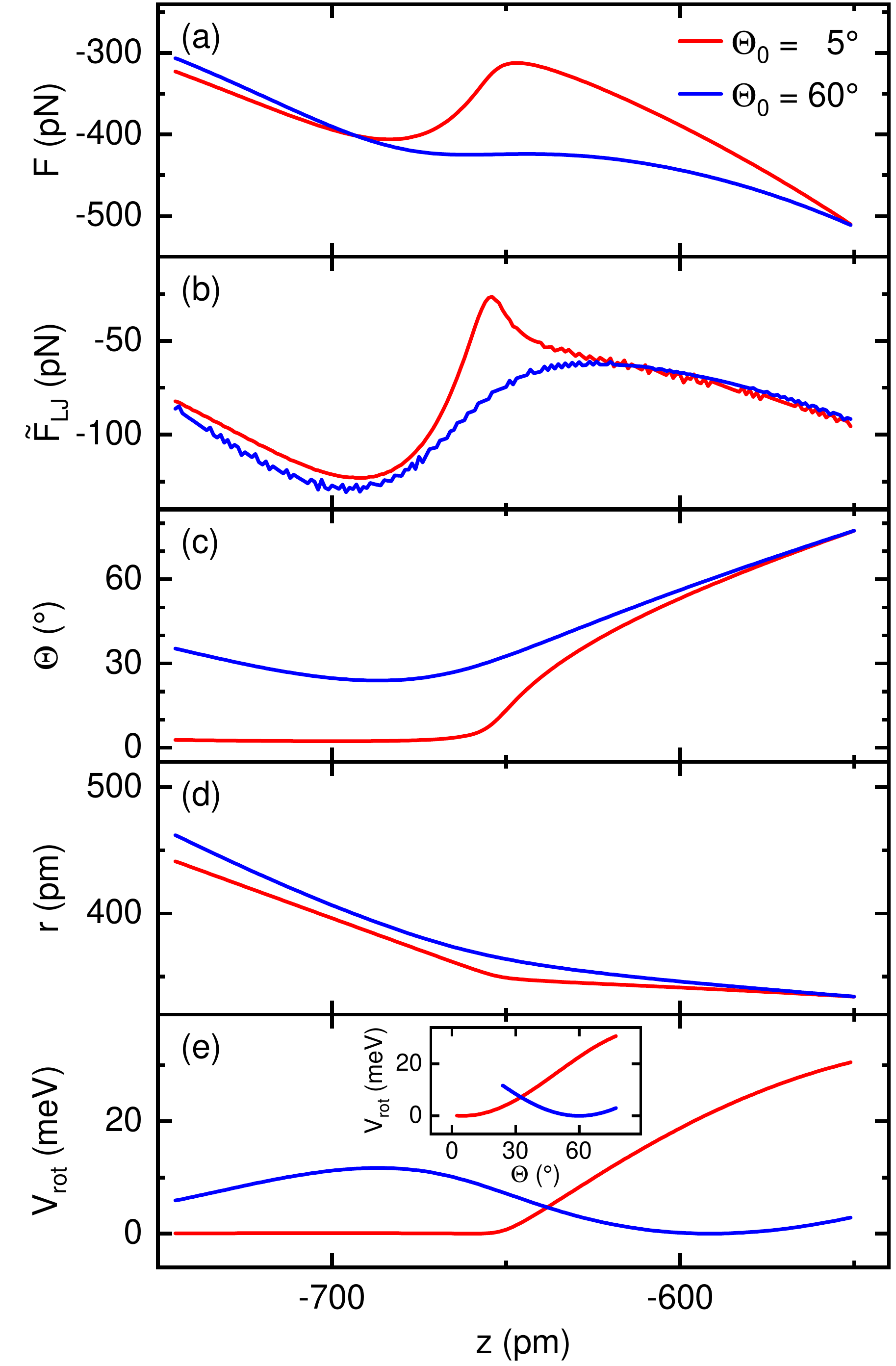}
\caption{
Calculated (a) total vertical force $F$ (eq \ref{eqS:F}), (b) Lennard-Jones force $\tilde{F}_{\text{LJ}}$ (eq \ref{eqS:Flj}), (c) tilt angle $\Theta$, (d) distance $r$ between O and Cu(111) and (e) $V_{\text{rot}}$ as a function of the tip displacement $z$ for initial tilt angles $\Theta_0=5^\circ$, $60^\circ$\@.}
\label{figS5}
\end{figure}

Figure \ref{figS5} shows in detail calculated results and the origin of the dip-hump evolution of the vertical force $F$\@.
For clarity, the discussion is restricted to a small and a large initial tilt angle, $\Theta_0=5^\circ$ and $\Theta_0=60^\circ$\@.
In Figure \ref{figS5}a the total vertical force resulting from
\begin{equation}
F=-\frac{\text{d}V}{\text{d}z}=-\frac{\text{d}}{\text{d}z}\left(V_{\text{rot}}+V_{\text{vdW}}+\tilde{V}_{\text{LJ}}\right)
\label{eqS:F}
\end{equation}
is plotted where $V_{\text{rot}}$, $V_{\text{vdW}}$, $\tilde{V}_{\text{LJ}}$ are defined in eq 1,2,4 of the article.
While for $\Theta_0=5^\circ$ a clear dip-hump evolution can be discerned, $\Theta_0=60^\circ$ yields an efficiently suppressed local maximum of $F$\@.
For all initial tilt angles the local minimum of $F$ reflects the contact of the O atom with the Cu(111) surface.
By comparing the evolution of $F(z)$ (Figure \ref{figS5}a) with 
\begin{equation}
\tilde{F}_{\text{LJ}}=-\frac{\text{d}\tilde{V}_{\text{LJ}}}{\text{d}z}
\label{eqS:Flj}
\end{equation}
 it becomes clear that the maximum of $F$ is influenced by the Pauli repulsion term of the modified Lennard-Jones potential $\tilde{V}_{\text{LJ}}$\@.
In addition, the long-range van der Waals attraction contributes to the decrease of $F$ with the reduction of the tip--surface distance.
This influence is due to the bending of the molecule (Figure \ref{figS5}c) because the distance between the O atom and the Cu(111) surface, $r$ (Figure \ref{figS5}d), decreases significantly more slowly than the displacement of the macroscopic tip, $z$.

The attenuation of the dip-hump structure, that is, the weakening of the local $F$ maximum, with increasing initial tilt angle $\Theta_0$ is best explained by exploring the evolution of $V_{\text{rot}}$ with $z$ (Figure \ref{figS5}e)\@.
After reaching the point of maximum attraction (Figure \ref{figS5}a) the CO molecule with large initial tilt angle bends back by a considerable amount under the influence of the Pauli repulsion towards its initial tilt angle.
Therefore, energy is gained in the harmonic potential $V_{\text{rot}}$ (Figure \ref{figS5}e), while $r$ and, thus, $\tilde{V}_{\text{LJ}}$, change weakly.
The inset to Figure \ref{figS5}e shows that for large $\Theta_0$ (right graph in the inset) the CO molecule adopts a substantial range of tilt angles $\Theta\leq\Theta_0$ during the tip approach to the surface.
When the repulsive Pauli force starts to exert a torque the molecule has to bend back to its initial tilt angle $\Theta_0$ and, therefore, reduces the energy in the harmonic potential $V_{\text{rot}}$\@.
In contrast, for small $\Theta_0$ (left graph in the inset) the Pauli torque tilts CO away from its equilibrium position, which costs energy due to the increase of $V_{\text{rot}}$\@.

\section{Comparison between Cu(111) and adsorbed Cu atom}

\begin{figure}[t]
\centering
\includegraphics[width=0.95\linewidth]{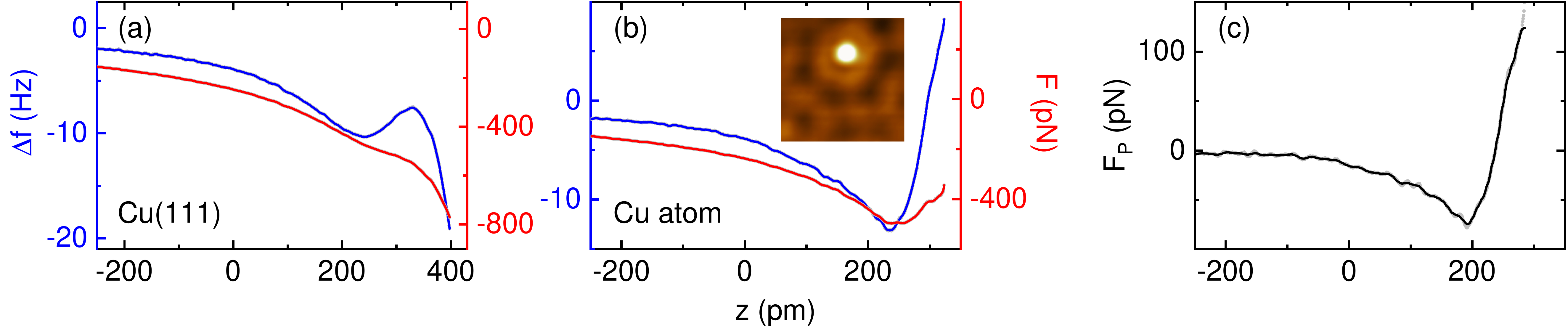}
\caption{ 
(a),(b) Frequency shift $\Delta f$ (blue) and force $F$ (red) acquired above the bare Cu(111) surface (a) and an adsorbed Cu atom (b) as a function of the displacement $z$ of the same CO-terminated tip.
Zero displacement corresponds to the tip--surface distance prior to disabling the feedback loop ($50\,\text{mV}$, $100\,\text{pA}$)\@.
Inset to (b): STM image of a single Cu atom adsorbed on Cu(111) ($50\,\text{mV}$, $100\,\text{pA}$, $5\,\text{nm}\times 5\,\text{nm}$)\@.
(c) Resulting short range force $F_\text{P}$ between the CO-terminated tip and the adsorbed Cu atom (see text).
}
\label{figS6}
\end{figure}

Figure \ref{figS6} compares the variation of $\Delta f$ and $F$ obtained for a CO-terminated tip on clean Cu(111) (Figure \ref{figS6}a) and on an adsorbed Cu atom (Figure \ref{figS6}b)\@.
The STM image (inset to Figure \ref{figS6}b) shows the adsorbed Cu atom as a circular protrusion on Cu(111)\@.
While the dip-hump structure is visible in the data sets obtained from the clean surface (Figure \ref{figS6}a), a Lennard-Jones-type behavior is observed for the adsorbed Cu atom (Figure \ref{figS6}b)\@.
Figure \ref{figS6}c shows the contribution of the short-range force, $F_\text{P}$, to the total vertical force, $F$\@.
It is obtained by subtracting the force on the clean Cu(111) surface, $F_\text{s}(z)$, from the force acquired above the adsorbed Cu atom, $F_\text{a}(z)$, i.\,e., $F_\text{P}(z)=F_\text{a}(z)-F_\text{s}(z)$\@. 
Different initial tip--surface distances were corrected by the apparent height difference between the Cu atom and the Cu(111) surface in STM data recorded with the CO-terminated tip.
A single minimum is observed around $\approx -70\,\text{pN}$, in agreement with a previous report \cite{science_366_235}.

\providecommand{\latin}[1]{#1}
\makeatletter
\providecommand{\doi}
  {\begingroup\let\do\@makeother\dospecials
  \catcode`\{=1 \catcode`\}=2 \doi@aux}
\providecommand{\doi@aux}[1]{\endgroup\texttt{#1}}
\makeatother
\providecommand*\mcitethebibliography{\thebibliography}
\csname @ifundefined\endcsname{endmcitethebibliography}
  {\let\endmcitethebibliography\endthebibliography}{}

\end{document}